\documentclass[12pt,3p,preprint,sort&compress]{elsarticle}
\usepackage[colorlinks=true]{hyperref}

\usepackage{amsmath}
\usepackage{amssymb}
\usepackage{mathtools}
\usepackage[utf8]{inputenc}

\newcommand{\eps}{\epsilon}

\renewcommand{\l}{\left(}
\renewcommand{\r}{\right)}

\newcommand{\be}{\begin{equation}}
\newcommand{\ee}{\end{equation}}
\newcommand{\ba}{\begin{align}}
\newcommand{\ea}{\end{align}}
\newcommand{\bg}{\begin{gather}}
\newcommand{\eg}{\end{gather}}
\newcommand{\bseq}{\begin{subequations}}
\newcommand{\eseq}{\end{subequations}}

\begin{document}
\begin{flushright}
	INR-TH-2022-016
\end{flushright}

\title{On direct observation of millicharged particles at $c$-$\tau$ factories and other $e^+e^-$-colliders} 
\author[inr,mpti]{Dmitry Gorbunov}
\ead{gorby@ms2.inr.ac.ru}
\author[inr,mpti]{Dmitry Kalashnikov}
\ead{kalashnikov.d@phystech.edu}
\author[lpi,hse]{Pavel Pakhlov}
\author[lpi,mpti]{Timofey Uglov}
\ead{uglov.timofey@gmail.com}
\address[inr]{Institute for Nuclear Research of Russian Academy of Sciences, 117312 Moscow, Russia}
\address[mpti]{Moscow Institute of Physics and Technology, 141700 Dolgoprudny, Russia}
\address[lpi]{P.N. Lebedev Physical Institute of the RAS, Moscow, Russia}
\address[hse]{Higher School of Economics (National Research University), Moscow, Russia}
\begin{abstract}
     Hypothetical particles with tiny electric charges (millicharged particles or MCPs) can be produced in electron-positron annihilation if kinematically allowed. Typical searches for them at $e^+e^-$ colliders exploit a signature of a single photon with missing energy carried away by the undetected MCP pair. We put forward an idea to look alternatively for MCP energy deposits inside a tracker, which is a direct observation. The new signature is relevant for non-relativistic MCPs, and we illustrate its power on the example of the $c$-$\tau$ factory, where we argued that the corresponding searches may be background-free. We find that it can probe the MCP charge down to $3\times10^{-3}$ of the electron charge for the MCP masses in ${\cal O}\l 5\r$\,MeV vicinity of each energy beam value where the factory will collect a luminosity of 100\,fb$^{-1}$ in one year. This mass region is unreachable with the searches for missing energy and single photon.  
\end{abstract}
\date{}

\maketitle

{\bf 1.} 
Feebly Interacting Massive Particles (FIMPs) is one of the options considered in various extensions of the Standard Model of particle physics (SM). The SM extensions are intended to explain the dark matter phenomena, baryon asymmetry of the Universe, neutrino oscillations, and other issues which the SM fails to describe. While the name FIMPs had been suggested for dark matter candidates with coupling to SM particles much weaker than that of WIMPs (weakly interacting massive particles), see e.g.\,\cite{Hall:2009bx,Bernal:2017kxu}, afterward it was used for non-dark matter candidates as well (and sometimes replaced by FIPs for Feebly Interacting Particles, see e.g.\,\cite{Agrawal:2021dbo}). A natural example of FIMPs is provided by so-called millicharged particles (MCPs)---hypothetical particles with tiny electric charge---which are predicted in SM extensions with vector portal coupling(s) to the hidden sector(s)\,\cite{Holdom:1985ag}, and are generically allowed with Abelian gauge subgroups popped up at high energies. 

The tiny electric charge provides stability to the MCPs and, hence, makes them a natural candidate for dark matter\,\cite{Goldberg:1986nk}. Anyway, if present, MCPs can impact on cosmology as a (main) part of dark matter sector\,\cite{Dubovsky:2003yn,Melchiorri:2007sq}, boost the star evolution and supernova explosion as additional source of cooling\,\,\cite{Dobroliubov:1989mr,Davidson:1993sj,Chang:2018rso}, induce the muon $g-2$ anomaly\,\cite{Bai:2021nai}, explain the EDGES signal at 21\,cm\,\cite{Liu:2019knx,Aboubrahim:2021ohe}, etc. Direct searches for MCPs have been performed in cosmic rays, at particle colliders, at beam dump experiments, and in experiments investigating neutrino physics, see e.g. Ref.\,\cite{Lanfranchi:2020crw} for a review and Fig.5.9 of Ref.\,\cite{Feng:2022inv} for the most recent summary of experimental constraints on the charge and mass of MCP. 

In this letter, we discuss MCP searches at $e^+e^-$ colliders, where they can be directly produced in pairs by a virtual photon thanks to their non-zero electric charge. 
Emerged MCPs having a small charge $\epsilon\ll1$ (in terms of the electron charge $e$) practically do not interact with the detector material and freely escape any detection. 
Since at $e^+e^-$ colliders initial 3-momenta are fixed, as a MCP signature the missing energy (taken away by MCPs) and a single photon (emitted by the colliding leptons) is generally adopted. This signature has been widely used in MCP searches at $e^+e^-$ colliders; the obtained limits are from e.g. LEP-I\,\cite{Davidson:1991si}, LEP-II\,\cite{Davidson:2000hf}, BESS-III\,\cite{Liu:2018jdi}, BaBar\,\cite{Liang:2019zkb}.  It will be also exploited in future searches at Belle II\,\cite{Liang:2019zkb} and developing projects like $\tau$-$c$ factory\,\cite{Liang:2019zkb}, $c$-$\tau$ factory\,\cite{Gorbunov:2022dgw} and CEPC\,\cite{Liu:2019ogn}. 

{\bf 2.} Here we propose another signature, which for $e^+e^-$ colliders, as we show below, turns out to be more sensitive to models with MCP (as compared to the traditional missing energy), but is applicable to rather limited regions in the model parameter space: namely, for the MCP mass $m_\chi$ close to the half of the $e^+e^-$ collision energy $\sqrt{s}/2\approx m_\chi$. In this case, despite the smallness of the electric charge, MCPs are non-relativistic and still able to produce hits in the tracker either by ionization or by knocked-on $\delta$-electrons. A hit probability is small, thus a full-ﬂedged track is unfeasible. However, for a certain range of the model parameters the MCP hit pattern is so distinctive that it will be possible to identify the process under the study without background as we demonstrate below.

Indeed, tracks of even non-relativistic MCPs, moving with velocity $\beta$ are moderately bent by the detector magnetic field: the Larmor radius in the magnetic field $B$ reads
\begin{equation}
\label{radius}
    r\approx 97\,\text{cm}\times \l\frac{m_\chi}{1\,\text{GeV}}\r\l\frac{1\,\text{T}}{B}\r \frac{\beta}{\epsilon}\,.
\end{equation}
For MCP to be recognized they must leave the inner tracker volume, so we require $r>20$ cm. 
Two MCPs are produced back-to-back, thus their hits must be on the specifically curved line crossing the interaction point. With sufficiently large number of hits the MCP tracks are clearly recognizable. 

Let us consider the propagation of nonrelativistic MCP in the medium of density $\rho$ and molar mass $A$, comprised of nuclei with electric charge $Z$. 
The mean number of initiating ionisation collisions for the MCP covered a  distance of $\delta x$ can be obtained from the Rutherford differential cross section (see e.g.\,\cite{ParticleDataGroup:2020ssz})
\begin{equation}
\label{loss_Rutherford}
    N(\delta x) = \frac{K}{2} \rho \frac{Z}{A} \frac{\epsilon^2}{\beta^2} \, \delta x \times \int_{I}^{T_{max}} \!\!\!\!\text{d} T \,\frac{1-\beta^2T/T_{max}}{T^2}\,, 
\end{equation}
where $I$ is the effective energy of ionization,  
 $K = 0.307 \times 10^6$\,eV\,cm$^2$\,mol$^{-1}$, and the maximal energy transfer from MCP to electron $T_{\text{max}}$ comes from the scattering kinematics as    
\begin{equation}
\label{Tmax}
    T_{\text{max}} = \frac{2m_e\beta^2\gamma^2}{1+\frac{2\gamma m_e}{m_\chi} + \left( \frac{m_e}{m_\chi} \right)^2}\,.
\end{equation}
In the limit of heavy and slow MCP, $m_\chi\gg m_e$, $\beta\ll 1$,  
one finds $T_{\text{max}}\to 2m_e\beta^2$ and 
\begin{equation}
\label{loss}
    N(\delta x) = \frac{K}{2} \rho \times \frac{Z}{A} \times \frac{\epsilon^2}{\beta^2} \, \delta x \times \left( \frac{1}{I} - \frac{1}{T_{max}} \right) \,. 
\end{equation}

To get the numerical estimate, we concentrate on the detector's drift 
chamber\,\cite{BASOK2021165490} suggested for the project of super $c$-$\tau$ factory designed in Budker Institute of Nuclear Physics (Novosibirsk), see many details at \texttt{https://sct.inp.nsk.su/}. 
The ionisation camera is supposed to be filled with propane, then $Z=26$, $A\approx 44$\,g\,mol$^{-1}$, $I\approx 11$\,eV and $\rho=1.864\times 10^{-3}$g\,cm$^{-3}$. 
Substituting these numbers to \eqref{loss} one arrives at 
\begin{equation}
\label{loss-number}
    N(\delta x) \approx 0.5 \times \l \frac{\delta x}{1\,\text{cm}}\r 
    \l \frac{\epsilon}{1.8\cdot 10^{-3}}\r^2 
    \l \frac{10^{-2}}{\beta}\r^2\,. 
\end{equation}
Hence, the small electric charge of MCP may actually be compensated by the small velocity to make the noticeable ionization as the ordinary charged particles, e.g. electrons, do. 

The tracker is a cylinder of $L=60$\,cm radius. The detector's drift chamber consists of $n_c=41$ sense wire layers. The basis of wire arrangement is hexagonal cell with typical edge length $\approx 1.46$\,cm. Introducing on average $\delta x=1$\,cm in Eq.\,\eqref{loss-number} we obtain the average number of ionisation events inside a cell 
\begin{equation}
 \label{loss-a} 
 N_1\equiv N(1\,\text{cm})\,.
\end{equation}
The number of cells lightened by the MCP pair along the straight trajectory inside the chamber is Poisson-distributed with mean $N_1$, hence the probability that not less than $n_0$ cells are lighten up is 
\begin{equation}
\label{prob}
    P_s(n_0,N_1)=\sum_{n=n_0}^{2n_c}  (e^{-{N}_1})^{2n_c-n} \times (1-e^{-N_{1}})^n \times 
\frac{2n_c!}{n!(2n_c-n)!} \,.
\end{equation}

For a slightly bent trajectory one may insert a correspondingly larger number $n_c$ in this formula, but we use  $n_c=41$  to be conservative. A cell may be hit by background events: soft but numerous photons emitted by the passing beam bunches, cosmic rays, etc. The estimates show that with a collider operating at the highest luminosity any single cell at any moment can be found being lightened with probability up to 5\%\,\cite{Belle2TDR}. Therefore, these combinatorial background events can mimic the same signature as MCP track with probability 
\begin{equation}
    P_b=\sum_{n=n_0}^{2n_c}  (1-0.05)^{2n_c-n} \times 0.05^n \times 
\frac{2n_c!}{n!(2n_c-n)!} \,.
\end{equation}
The false track begins from a single cell which may be lightened by any sufficiently energetic photon passing through the drift chamber. These photons are generated at each bunch crossing. About $10^{16}$ bunch crossings are expected each year when the collider operates at the highest luminosity 1/ab/year\,\cite{Shekhtman_2020}. Zero background $10^{16}\times P_b(n_0)\ll 1$ implies $n_0=29$ and bigger. To be on the safe side, we put $n_0=30$ for the numerical analysis. 

A pair of produced at the collider MCPs must hit the same number of cells, $n_0=30$, or more to make the background-free signature, and its probability is $P_s(n_0,N_1)$. In what follows we calculate the number of produced MCP pairs as a function of MCP mass, charge and velocity, multiply it by this probability, which through Eqs.\,\eqref{loss-a}, \eqref{loss-number} depends additionally on MCP velocity and charge, and ask for the number of the evaluated in this way MCP tracks to exceed 3. That corresponds to the upper limit on the MCP charge to be placed at 95\% CL within the Poisson statistics. Indeed, this signature of the nonrelativistic MCP is background-free.

If MCP is sufficiently energetic, on its way through the chamber it can kick off an electron from an atom.  
The differential energy spectrum of these so-called $\delta$-electrons produced per unit length d$x$ by a travelling MCP of velocity $\beta$ reads\,\cite{ParticleDataGroup:2020ssz} 
\begin{equation}
\label{delta}
\frac{{\text d}^2N_\delta}{{\text d}x\,{\text d}T_e}=\rho\,\frac{K}{2} \frac{Z}{A} \frac{\epsilon^2}{\beta^2}  \frac{F(T_e)}{T_e^2}\,,
\end{equation}
where for non-relativistic electrons one can set $F(T_e)\approx 1$. The estimate \eqref{delta} is valid for MCP faster than electrons bound in atom, i.e. $\beta>\alpha$, and for sufficiently energetic MCPs. In practice, the electron energy must exceed some value, e.g.  
\begin{equation}
\label{Tmin}
    T_e>T_{\text{min}}=1\,\text{keV}\,,
\end{equation}
to be detected through the subsequent ionization. Thus the number of viable $\delta$-electrons, which are produced by MCP passed a distance $L$, can be estimated as 
\begin{equation}
\label{N-delta-e}
N_\delta=\rho\,\frac{K}{2} \frac{Z}{A} \frac{\epsilon^2}{\beta^2} L \l \frac{1}{T_{\text{min}}}-\frac{1}{T_e}\r\,.
\end{equation}
Recall that the kinematics limits the maximum energy transfer from MCP to the $\delta$-electron as \eqref{Tmax}. Since MCP must be at least moderately nonrelativistic, given Eq.\,\eqref{N-delta-e} and \eqref{Tmin} there is only a narrow interval of $0.03\leq \beta\ll 1$ where $\delta$-electrons may be emitted. This process can be treated as a supplemental to the ionisation (rather than  standalone) signature.

{\bf 3.} 
So we look at the direct production of MCPs close to their threshold, 
\[
e^++e^-\to \chi+\bar\chi\,,
\]
where they are essentially non-relativistic, with velocity 
\[
\beta\equiv \sqrt{\frac{\sqrt{s}}{m_\chi}-2}\ll 1.
\]
Hence their total production cross section is given by 
\begin{equation}
\label{ee_to_chi_chibar_X_section}
    \sigma_{e^+e^- \to \chi \bar{\chi}}(s) = \frac{2\pi\alpha^2\epsilon^2\beta}{s}\,,
\end{equation}
which can be rescaled from that of muon pair production at the threshold. 

Since the center of mass energy is not exactly monochromatic due to a beam energy smear, we approximate it by the Gaussian shape with mean value $\sqrt{s_0}$ and dispersion $\sigma_0^2$. If the collider operates for some time at a given energy $\sqrt{s_0}$ and collects an integrated luminosity ${\cal L}_0$, the effective differential luminosity is 
\begin{equation}
    \text{d}L = \frac{{\cal L}_0}{\sqrt{2\pi} \sigma_0} \times \text{exp}\left( -\frac{(\sqrt{s}-\sqrt{s_0})^2}{2\,\sigma_0^2} \right) \text{d}\sqrt{s}
\end{equation}
and the total number of the produced at this operation period MCP pairs can be estimated as 
\begin{equation}
\label{events}
    N (\sqrt{s_0}) = \int\!\! \text{d}\sqrt{s}\; \frac{\text{d}L}{\text{d}\sqrt{s}} \; \sigma _{e^+e^- \to \chi \bar{\chi}}(s)\,.
\end{equation}

The beam energy dispersion significantly reduces the production of non-relativistic MCP of a particular mass but instead extends the MCP production inside a mass range $m_\chi=\sqrt{s_0}/2\pm\sigma_0$. Initial State Radiation (ISR) leads to a similar effect, but as we show in due course, the ISR contribution in comparison with the beam energy smearing can be neglected in this respect. 

Finally, each MCP pair has a particular chance to ionise the tracker material sufficiently to be recognizable as we explained above. The chance depends on the MCP velocity.  Therefore the number of MCP-initiated tracks consisted of not less than $n_0=30$ lightened cells equals 
\begin{equation}
\label{tracks}
    N_t (\sqrt{s_0}) = \int\!\! \text{d}\sqrt{s}\; \frac{\text{d}L}{\text{d}\sqrt{s}} \; \sigma _{e^+e^- \to \chi \bar{\chi}}(s)\times P_s(n_0,N_1)\,.
\end{equation}
where $P_s$ and $N_1$ follow from Eqs.\,\eqref{prob} and \eqref{loss-number}. The number of expected $\delta$-electrons from each propagating MCP is given by 
\eqref{N-delta-e}. 

Another source of MCPs can be the decay of a hadronic vector resonance, which is produced at the threshold. It can directly decay into a nonrelativistic MCP pair. We find that among the resonances planned to be studied at the factory the only promising is $J/\psi$. The number of signal events from this source is
\begin{equation}
\label{Meson}
   N = N_{J/\psi} \times \eps^2 \times Br_{\chi \bar{\chi}} \times P_s (n_0,N_1) \,,
\end{equation}
where the number of produced mesons is $N_{J/\psi} = 5 \cdot 10^{11}$ and the meson branching into MCP $Br_{\chi \bar{\chi}}$ can be expressed via its branching 
into muons $Br_{\mu\mu}$ as follows \cite{Aloni:2017eny}
\begin{equation}
\label{Branching}
  Br_{\chi \bar{\chi}} = Br_{\mu\mu} \times \frac{\sqrt{1-4m_\chi^2/m_{J/\psi}^2} \l 1+2m_\chi^2/m_{J/\psi}^2\r}{\sqrt{1-4m_\mu^2/m_{J/\psi}^2} \l 1+2m_\mu^2/m_{J/\psi}^2\r} \,.
\end{equation}


{\bf 4.}  
To illustrate the efficiency of the suggested method in searches for MCP at $e^+e^-$ colliders we perform a numerical estimate for a tracker filled with propane and choosing  $\sqrt{s_0}=3$\,GeV, ${\cal L}_0=100$\,fb$^{-1}$, $\sigma_0=\sqrt{2} \times 0.1 \% \times \sqrt{s_0}/2$ and so $\sigma_0$ equals 2.1\,MeV. We assume that the ionization signature is reliable for identification of MCP pair if the number of ionised cells definitely exceeds $n_0=30$, and hence each MCP at least covers a distance of about $X_{min}=15$\,cm. 
The ionisation intensity must be reasonably high, e.g. $N_1\gtrsim 0.13$, see \eqref{loss-number}.

With the chosen collider and detector parameters, one can roughly limit the attainable MCP velocities from below. Indeed, for the ionization the lowest recognizable intensity \eqref{loss-number} $N_1\gtrsim 0.13$ with the number of hits exceeding $n_0=30$ and the lowest energy transfer $I$ imply $\beta>5\times 10^{-3}$. The signature with $\delta$-electrons for \eqref{Tmax} and \eqref{Tmin} to be fulfilled requires $\beta>3\times 10^{-2}$. We observe below that the relevant velocities always exceed $1\times 10^{-2}$, which is still within the borders of applicability of the energy loss formulas we use, \eqref{loss}, \eqref{delta}. 

The numerical analysis is accomplished by scanning over $m_\chi$ and $\epsilon$, integrating over the beam energy spread in \eqref{tracks} while checking all the constraints on the velocity. To estimate the sensitivity of these searches to the MCP exhibiting the ionisation signature, we compare the result with 3 corresponding to 95\% CL within the Poisson statistics. If the number exceeds this value, the region around this particular point in ($m_\chi$,$\epsilon)$ plane can be tested with our method.   

\begin{figure}[!htb]
    \centering
    \includegraphics[width=0.8\textwidth]{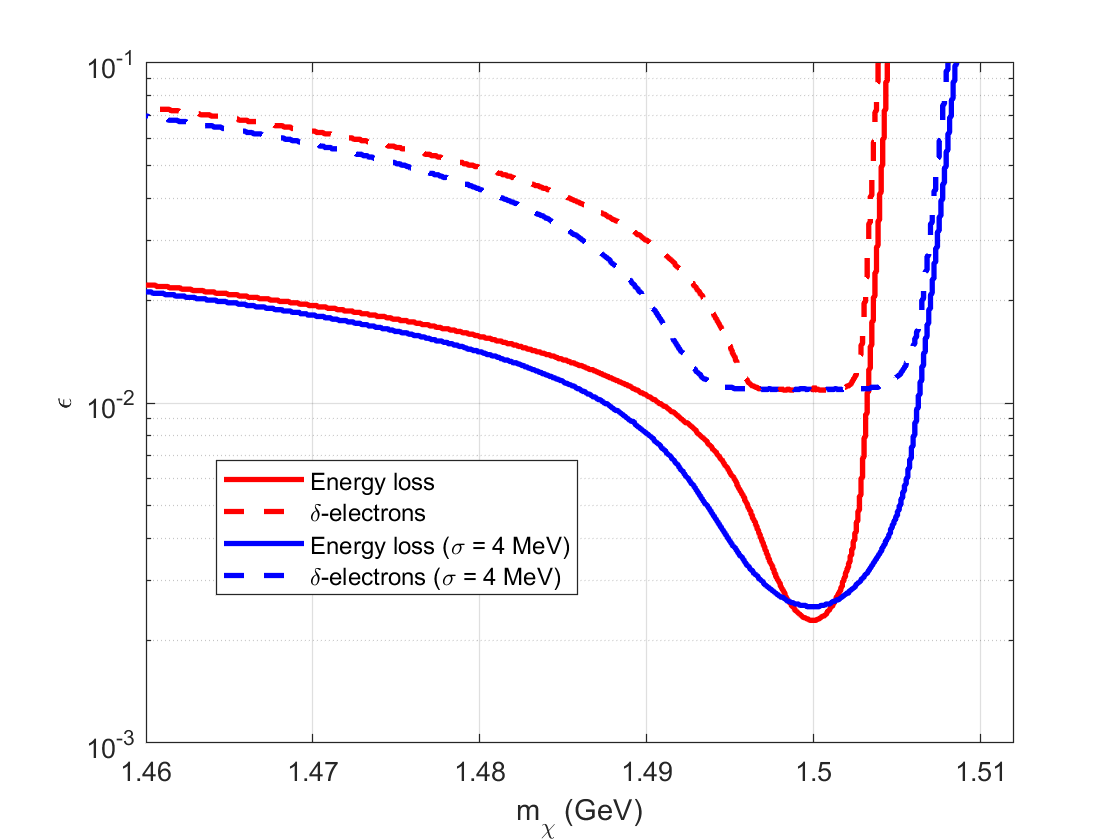}
    \caption{The outlined regions would provide with 3 or more signal events upon collecting ${\cal L}_0=100$\,fb$^{-1}$ integrated luminosity at collision energy $\sqrt{s_0}=3$\,GeV, see the main text for the detector parameters. The red lines are obtained with beam energy spread $\sigma_0=\sqrt{2} \times 10^{-3}\times \sqrt{s_0}/2=2.1$\,MeV and the blue lines with $\sigma_0=4$\,MeV.}
    \label{fig:fixed}
\end{figure}

To estimate approximate values of $\epsilon$ and $\beta$ and to illustrate the effect of the beam energy smear we carried out simplified calculations where instead of using probability (\ref{prob}) we put limits on $\epsilon/\beta$ from the requirement on ionisation energy loss $dE/dx>25$ eV/cm. It corresponds to $N_1 \approx 0.5$. The numerical results are presented in Fig.\,\ref{fig:fixed}. 
In the model with parameters in the upper regions outlined by red solid (dashed) lines one expects more than 3 MCP events with ionization ($\delta$-electron) signature described above, and the upper region will be excluded at 95\% CL if no events are observed. The similar blue lines show the same limits but obtained with larger beam energy spread, $\sigma_0=4$\,MeV. Evidently with a broader beam energy one probes a wider MCP mass range, but the probing is poorer at the very threshold for the ionization and remains the same for $\delta$-electrons. The latter is because the lower limit on $\epsilon$ here is related to the minimal $\beta$ from \eqref{Tmin} via \eqref{N-delta-e}. The range of MCP signature with $\delta$-electron is limited from above by kinematics related to $T_{min}$ and $\sigma_0$. The similar kinematics works in case of ionization, but the minimal possible velocity here is lower, and so the smaller charges are accessible. The width of the mass region where the searches exhibit the highest sensitivity is the same for both signatures, and it is determined by the beam width. The position of each region is defined by kinematics, that is the minimal accessible kinetic energy or the minimal velocity, which is somewhat different. The sensitivity to lighter MCPs are limited by the intensity limit \eqref{loss} and minimal length $X_{min}$ for the ionization signature and the number of kicked electrons \eqref{N-delta-e} for $\delta$-electrons. 

We observe that a higher sensitivity to MCPs are exhibited by the ionization signature. The region where the $\delta$-electron signature may show up is always embedded into the region where the ionisation signature works. 

At first glance, both signatures we discuss work perfectly as far as $\epsilon\sim\beta$, see eqs.\,\eqref{loss} and \eqref{delta}. Inserting this relation into the cross section\,\eqref{ee_to_chi_chibar_X_section} one finds that the requirement to have a few events with statistics of 100\,fb$^{-1}$ would imply testing the charge as low as $10^{-3}$, while our numerical calculation reveals 3 times larger value. This discrepancy is due to the non-monochromaticity of the colliding beams. Indeed, the velocity $10^{-3}$ implies the MCP kinetic energy of 750\,eV, while 
the 0.1\% beam energy spread we use in our estimates implies the mass range of about 1.5\,MeV. Hence, the production of MCP with kinetic energy below 750\,eV is highly suppressed because of much lower effective luminosity. Since the number of events in the monochromatic case scales as $\beta^3$, and the MCP kinetic energy scales as $\beta^2$, the typical velocity of the produced MCP is by a factor $(1.5\times 10^3/0.75)^{1/5}\approx 5$ higher than our naive estimate $10^{-3}$, that nicely fits to the numerical results shown in Fig.\,\ref{fig:fixed}.       

\begin{figure}[!htb]
    \centering
    \includegraphics[width=0.8\textwidth]{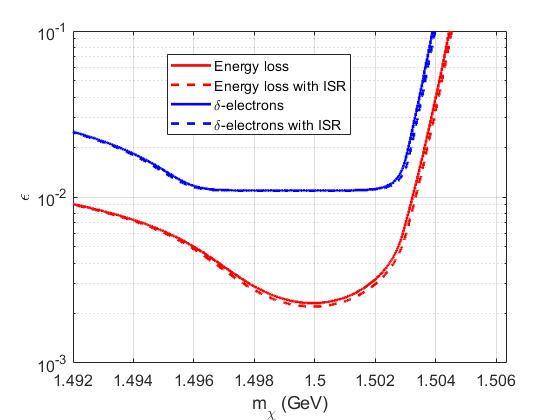}
    \caption{The outlined regions would provide with 3 or more signal events upon collecting ${\cal L}_0=100$\,fb$^{-1}$ integrated luminosity at collision energy $\sqrt{s_0}=3$\,GeV, with $\sigma_0=2.1$\,MeV,  see the main text for detector parameters. The solid lines are obtained without contribution of ISR and the dashed lines with ISR.}
    \label{fig:ISR}
\end{figure}

It is worth to note that the center-of-mass energy of the colliding $e^+e^-$ pair must be corrected for the unavoidable Initial State Radiation, which escapes detection. 
This can be done with the following extension of eq.\,\eqref{events},  
\begin{equation}
\label{eventsE}
    N (\sqrt{s_0}) = \int\!\! \text{d}\sqrt{s}\; \int_0^{x_{max}} \text{d}x \;\frac{\text{d}L}{\text{d}\sqrt{s}} \; \sigma _{e^+e^- \to \chi \bar{\chi}}((1-x)s) \, H(x,s) \,,
\end{equation}
where the kernel $H(x,s)$ is presented in Ref.\,\cite{Kuraev:1985hb} and $x$ refers to the energy fraction carried away by the emitted (and missed) photons as $x=1-s'/s$; the maximal fraction is taken to be 
$x_{max} = (25 \, \text{MeV})^2/s$. 
We find that for the interesting energy and mass ranges the kernel can be approximated simply as 
\[
 H(x,s) = B (x^{B-1}-1+x/2)\,,\hskip 1cm \text{with} \hskip 1cm  B=\frac{2\alpha}{\pi} \l \log\l \frac{s}{m_e^2} \r -1 \r. 
\]
We repeat the previous analysis, carrying out simplified calculations, and obtain the limits presented in Fig.\,\ref{fig:ISR}. 
One observes that the corrections due to ISR are small and we neglect them in what follows. 

The estimated sensitivity above is obtained upon zero background conditions given the suggested signatures. Then, since the number of signal events scales as $N\propto {\cal L}\times \epsilon^2\times \epsilon^2$, with ten times higher statistics, 1\,ab$^{-1}$, the overall sensitivity will increase by a factor of $1.8$. The sensitivity of the ionization technique can be extended to smaller masses with lower $X_{min}$ and to bigger and smaller masses and even to smaller $\epsilon$ with lower $I$. Naturally the sensitivity of our signatures will be also improved with denser gas in the tracker, but this way is limited as it alters the zero background conditions. 

\begin{figure}[!htb]  
\centerline{
\includegraphics[width=0.8\textwidth]{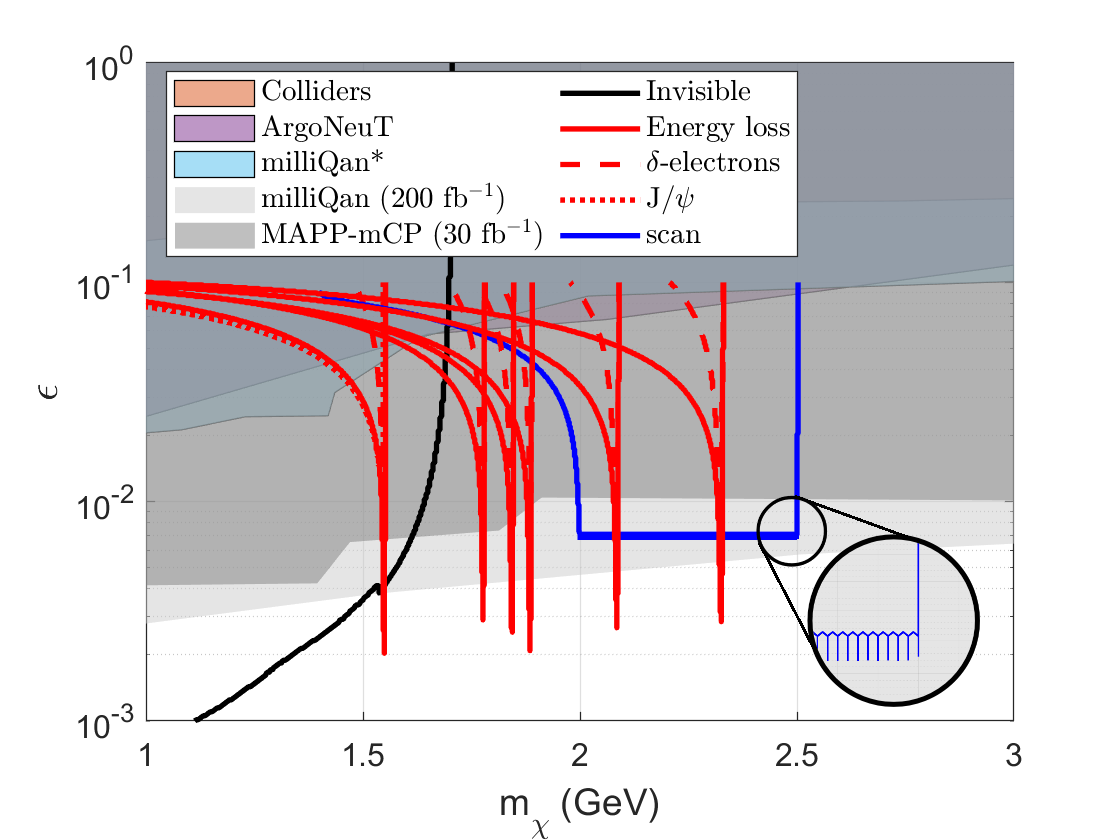} 
}
\caption{\label{fig:summary}
The regions to be probed at 95\%\,CL with $e^+ e^- \rightarrow \chi \chi$ and non-relativistic MCP signatures (red solid line for the ionisation and red dashed line for the $\delta$-electrons) in the tracker after one year of operation following the scientific program of $c$-$\tau$ factory {\texttt{https://sct.inp.nsk.su}}, see Tab.\,\ref{tab1}. The smallest charge to be tested at different thresholds is somewhat different because of somewhat different luminosity there. The total collision statistics for one year is 1000\,fb$^{-1}$. The region above the solid blue line ($\epsilon\approx 7\times 10^{-3}$ with 5.7\% vertical jumps) can be investigated with the scanning over the energy as explained in the main text following of the plan of $\tau$-$c$ factory\,\cite{Achasov:2023gey}.  
The shaded outlined regions are excluded at 95\% CL with searches undertaken by collaborations ArgoNeuT\,\cite{ArgoNeuT:2019ckq}, milliQan\,\cite{Ball:2020dnx}, and at terminated  colliders\,\cite{Davidson:2000hf}. The shaded in grey regions will be explored at future experiments MAPP-mCP and milliQan\,\cite{Acharya:2022nik}, the region above solid black line can be probed with standard missing energy searches\,\cite{Gorbunov:2022dgw}
at the $c$-$\tau$ factory with the same luminosity and the same operation schedule as in Tab.\,\ref{tab1}.} 
\end{figure}

Typically an $e^+e^-$ machine either operates at several specially chosen collision energies $\sqrt{s_0}$ or performs a scan over some energy range. This program would allow to probe a broader region of MCP masses. To illustrate the realistic prospects we assume that the $c$-$\tau$ factory follows the plan to operate for one year at several energies near the thresholds of interesting hadronic resonances, see Tab.1.1 in the CDR (part one) of the proposal at {\texttt{https://sct.inp.nsk.su}}, which we retype in Tab.\ref{tab1}, 
\begin{table}[!htb]
    \centering
    \begin{tabular}{|c|c|c|c|c|c|c|}
    \hline
    $\sqrt{s}$\,,GeV & $3.097$ & $3.554$ & $3.686$ & $3.770$ & $4.170$ & $4.650$ \\
    \hline
    ${\cal L}_0, \, \text{fb}^{-1}$ & $300$ & $50$ & $150$ & $300$ & $100$ & $100$ \\
    \hline
    \end{tabular}
    \caption{Energies, at which Super Charm--Tau factory data will operate and the integrated luminosity over these energies.}
    \label{tab1}
\end{table}
The total collected luminosity is  1000\,fb$^{-1}$. To investigate the opportunities of 
$c$-$\tau$ factory with this program we carried out the complete analysis using probability (\ref{prob}). The results are depicted in Fig.\,\ref{fig:summary}.  
There we also shade the regions excluded at 95\% CL by previous searches with particle accelerators. 
In this respect it is worth to mention other competitive limits from analysis\,\cite{Plestid:2020kdm} of Super-K results and reinterpretation\,\cite{Marocco:2020dqu} of the results of BEBC WA66 experiment.  

To illustrate what the scanning over the beam energy can achieve we follow the schedule presented in Tab.\,2.1 of Ref.\,\cite{Achasov:2023gey} with CDR of the China's project of Super $\tau$-c factory, where the scan step is 10\,MeV and at each position of the collision energy an integrated luminosity 1\,fb$^{-1}$ is planned to be collected. For the illustration purpose we chose the beam energy range $\sqrt{s}/2=$2-2.5\,GeV and at each of the 101 positions use the same track signature in the ionisation camera, which will allow for exploring the regions in the model parameter space above the solid blue line depicted in Fig.\,\ref{fig:summary}.

{\bf 5.} To conclude, we propound a new method to search for hypothetical millicharged particles produced in electron-positron collisions. It exploits a high ionization power of non-relativistic particles, which compensates for the smallness of their electric charge. The method enables to probe of the regions of masses close to the threshold.  
We illustrate the perspectives of this method by applying it in the framework of proposed $c$-$\tau$-factory 
{\texttt{https://sct.inp.nsk.su}}. With the one-year program of collecting data at certain collision energies (related to the thresholds of interesting particles) presented in Tab.\,\ref{tab1} 
one can be able to test the charges as small as $\epsilon\sim 3\times 10^{-3}$. The scanning over the collision energy ($\sqrt{s},\sqrt{s}+\Delta$) with 10\,MeV step and integrated luminosity of 1\,fb$^{-1}$ per step, proposed for the super $\tau$-$c$ factory\,\cite{Achasov:2023gey}, will give a possibility to explore the MCP charges down to $7\times 10^{-3}$ over the whole region of of MCP masses ($\sqrt{s}/2,\sqrt{s}/2+\Delta/2$).

The sensitivity can be increased with narrower beam energy of colliding particles, though it would shrink the mass range under investigation. Since the number of signal events scales as $\epsilon^4$, an increase in sensitivity by a factor of 2 requires 16 times longer operation period. In this respect the "standard" technique of searches for missing energy (carried away by MCPs), where the number of events scales as $\epsilon^2$ is more promising, though it suffers from (ir)reducible background (see e.g. study of missing energy signature at $\tau$-$c$ factory\,\cite{Zhang:2019wnz} and at $c$-$\tau$ factory \cite{Gorbunov:2022dgw}), which is absent in our case. As one observes from Fig.\,\ref{fig:summary}, the missing energy signature is not applicable for heavy MCP, which can be explored with ionisation as we discuss in this paper. 
The most attractive feature of our method is that we use the appearance as the signature, which immediately reveals the identity of new particles produced in collisions, while the missing energy is blind to its cause.  

In cases where $\beta/\epsilon$ is not very large the MCP Larmor radius\,\eqref{radius} is of the order of the drift chamber size. Therefore, a deviation of the trajectory from the straight line is expected, the trajectory radius can be measured that yields the information about MCP mass, charge and velocity. Similar information follows from measurements of the time of flight provided by the drift chamber registration system. Considering realistic non-monochromatic beams, the additional information can provide with better precision in MCP mass and charge inferred in this analysis.

\vskip 0.3cm
We thank A.\,Bondar, S.\,Demidov, S.\,Gninenko, I.\,Logashenko and K.\,Todyshev for valuable discussions.  The work on the expected signal coming from ionization energy loss of millicharged particles is supported by the Russian Science Foundation RSF grant 21-12-00379. The work of D.\,K. on the expected signal coming from $\delta$-electrons and $J/\psi$ is supported by the Foundation for the Advancement of Theoretical Physics and Mathematics “BASIS”. 

\bibliographystyle{utphys}
\bibliography{refs}
\end{document}